\documentclass{emulateapj}

\newcommand{\psiz}{0.8}
\newcommand{\psizf}{0.8}

\slugcomment{2010, ApJ, 714, 989}

\shorttitle{Spin Evolution: Effect of Magnetic Star-Disk Coupling}
\shortauthors{Matt et al.}

\begin{document}



\title{Spin Evolution of Accreting Young Stars. I. Effect of Magnetic Star-Disk Coupling}












\author{Sean P. Matt$^{1,2}$,
        Giovanni Pinz\'on$^{3,4}$,
        Ramiro de la Reza$^{5}$, and
        Thomas P. Greene$^{1}$}

\affil{$^1$NASA Ames Research Center, M.S. 245-6, Moffett Field, CA
  94035-1000, USA; sean.p.matt@nasa.gov, thomas.p.greene@nasa.gov}

\affil{$^2$NASA Postdoctoral Program Fellow}

\affil{$^3$Observatorio Astron\'omico Nacional, Facultad de Ciencias,
  Universidad Nacional de Colombia, Bogot\'a, Colombia;
  gapinzone@unal.edu.co}

\affil{$^4$Direcci\'on Nacional de Investigaciones, Universidad
  Antonio Nari\~no, Bogot\'a, Colombia}

\affil{$^5$Observat\'orio Nacional, Rio de Janeiro, Brazil;
  delareza@on.br}

\begin{abstract}

We present a model for the rotational evolution of a young, solar mass
star interacting with an accretion disk. The model incorporates a
description of the angular momentum transfer between the star and disk
due to a magnetic connection, and includes changes in the star's mass
and radius and a decreasing accretion rate.  The model also includes,
for the first time in a spin evolution model, the opening of the
stellar magnetic field lines, as expected to arise from twisting via
star-disk differential rotation.  In order to isolate the effect that
this has on the star-disk interaction torques, we neglect the
influence of torques that may arise from open field regions connected
to the star or disk.  For a range of magnetic field strengths,
accretion rates, and initial spin rates, we compute the stellar spin
rates of pre-main-sequence stars as they evolve on the Hayashi track
to an age of 3~Myr.  How much the field opening affects the spin
depends on the strength of the coupling of the magnetic field to the
disk.  For the relatively strong coupling (i.e., high magnetic
Reynolds number) expected in real systems, all models predict spin
periods of less than $\sim3$ days, in the age range of 1--3~Myr.
Furthermore, these systems typically do not reach an equilibrium spin
rate within 3~Myr, so that the spin at any given time depends upon the
choice of initial spin rate.  This corroborates earlier suggestions
that, in order to explain the full range of observed rotation periods
of approximately $1$--$10$ days, additional processes, such as the
angular momentum loss from powerful stellar winds, are necessary.

\end{abstract}



\section{Introduction} \label{sec_intro}

The last three decades of observational surveys of pre-main-sequence
stars have resulted in the measurement of rotation periods and/or
rotational velocities ($v \sin i$) of thousands of stars \citep[see
  reviews by][]{herbstea07, scholz09}.  The analysis of these data
have revealed a number of mysteries regarding the spin rates of young
stars.  Perhaps the most prominent of these mysteries is that which
was first noted by \citet{vogelkuhi81}.  Specifically, a large
fraction of nearly solar mass pre-main-sequence stars rotate much more
slowly than expected.  The fact that stars are built up from material
with high specific angular momentum, and that these young stars are
still contracting, leads to the expectation of a rotation rate near
the breakup velocity.

Solar mass stars with ages less than a few Myr have rotation periods
typically in the range of 1--10 days but with a tail in the
distribution extending to around 20 days
\citep[e.g.,][]{attridgeherbst92, choiherbst96, stassunea99, rebull01,
  rebull3ea04, coveyea05, scholz09}.  The statistics show that about
half of these stars are indeed fairly rapid rotators and do seem to
spin up as they approach the main sequence \citep{bouvier3ea97,
  delarezapinzon04, scholzea07, irwinea08}.  However, roughly half of
the stars younger than a few Myr rotate at about 10\% or less of
breakup speed \citep[][]{rebull3ea04, herbstea07, scholz09}.  Thus,
there is some mechanism operating that is capable of removing
significant amounts of angular momentum during the pre-main-sequence
phase.

Following the suggestion of \citet{edwardsea93}, the modeling work of
\citet{bouvier3ea97} and \citet{rebull3ea04} showed that the observed
range of spins could be reproduced reasonably well, by assuming that
the presence of an accretion disk somehow results in a constant
stellar spin period of around one week.  There is some observational
evidence that the population of stars with disks on average rotate
more slowly than those without disks \citep[e.g.,][]{edwardsea93,
  choiherbst96, stassunea99, herbstea00, littlefairea05, rebullea06,
  ciezabaliber07}.  Also, the idea that the angular spin rate is held
at a constant by the presence of the disk is loosly based on physical
models for the interaction between a star and surrounding disk.

There are generally two prominent theoretical ideas for how accreting
stars can maintain a slow spin rate.  One idea is that the torques
arising from the magentic connection between the star and disk can
remove substantial angular momentum \citep[e.g.][]{ghoshlamb78,
  camenzind90, konigl91, shuea94}.  When these torques are strong
enough to enforce an equilibrium stellar spin rate, this idea is
generally referred to as ``disk locking'' \citep{choiherbst96}.  The
other idea to explain slowly rotating accretors is that powerful
stellar winds are primarily responsible for removing angular momentum
from the star \citep{hartmannmacgregor82, mestel84,
  hartmannstauffer89, toutpringle92, paatzcamenzind96, mattpudritz05l,
  mattpudritz08II, mattpudritz08III}.  Numerical, dynamical
simulations of the star-disk interaction \citep[such as the
  investigations of spin equilibrium by][]{romanovaea02, long3ea05}
typically exhibit significant torques arising both from the magnetic
star-disk connection and from winds.

The goal of the present paper is to further examine the possibility of
disk locking as an explanation for the slow rotation of young stars.
To this end, we develop a model for the time-evolution of stellar spin
rates, under the influence of magnetic star-disk interaction torques.
In this work, we neglect the influence of stellar winds, as is
customary in the disk-locking models.

There are generally two types of disk locking models. One is the
X-wind model \citep[e.g.,][]{shuea94, ostrikershu95, mohantyshu08}.
This model is developed under the assumption that the star-disk system
on average exists in a spin equilibrium state, in which the accreting
star feels no torque.  Thus, the X-wind model cannot be used to
address whether or how the system reaches this equilibrium, nor how
the system behaves when out of equilibrium (e.g., due to variability
or due to the evolution of the star or disk).  All other disk locking
models \citep[e.g.,][]{camenzind90, konigl91, wang95, mattpudritz04}
are of the second type, based on the general picture developed by
\citet{ghoshlamb78} for accreting neutron stars.  The Ghosh \& Lamb
model provides a method for calculating the torque on the star, due to
the star-disk interaction, for any state of the system.  The star may
spin up or down, depending on conditions, and there is a hypothetical
disk-locked state, in which the net torque on the star is zero.  Thus,
the Ghosh \& Lamb model can be used to compute the evolution of
accreting star spins and to assess how far from equilibrium a system
may be.

\citet[][hereafter CC93]{cameroncampbell93}, \citet{yi94, yi95}, and
\citet[][hereafter AC96]{armitageclarke96} computed the evolution of
pre-main-sequence star spins, adopting a Ghosh \& Lamb-type model for
torques on the star, and also including stellar contraction, a
decrease in accretion rate with time, and a prescription for the
evolution of the magnetic field.  For parameters that reasonably
represent T Tauri stars, their models produced spin rates within the
observed range.  These models also showed that, regardless of the
initial spin of the star, the spin rates rapidly (in less than $\sim
1$~Myr) approached an equilibrium value and stayed near equilibrium
during the first few Myr of evolution.  That is, the initial spin
conditions were ``erased,'' and the stellar spin was ``locked'' to a
rate that depended only on the stellar mass and radius, magnetic field
strength, and accretion rate.  These results generally supported the
idea that disk locking explains the slow rotation of accreting T Tauri
stars.

However, since that work, it has been recognized that there is a
serious theoretical problem with a key assumption of the classical
Ghosh \& Lamb torque model.  The assumption is that the stellar
magnetic field lines connect to a large region of the disk and become
highly twisted in the azimuthal direction.  This twisting is what
gives rise to the magnetic spin-down torque felt by the star.  Several
authors \citep[e.g.,][]{vanballegooijen94, lyndenbellboily94,
  lovelace3ea95, hayashi3ea96, millerstone97, goodson3ea97,
  fendtelstner00, mattea02, uzdensky3ea02, romanovaea02, kuker3ea03,
  bessolazea08} have now shown that, when the dipole field is twisted
azimuthally past a threshold of approximately $45^\circ$, the magnetic
pressure associated with the azimuthal component pushes the field
outward.  This leads to an inflation and opening of the field loops in
a stellar rotation timescale, ultimately disconnecting the star from
the disk.  \citet{uzdensky3ea02} developed a method for computing
which magnetic field lines would open.  The amount of field opening
depends primarily on how strongly coupled the magnetic field is to the
disk, since the coupling is responsible for the twisting, and the
twisting is responsible for the field opening.

In order to assess how the field opening affects the torques,
\citet[][hereafter MP05]{mattpudritz05} modified the Ghosh \& Lamb
formulation to include the effects of field opening, following the
method of \citet{uzdensky3ea02}.  MP05 showed that, for the relatively
strong coupling expected in these systems, the opening of the field
significantly reduces the spin-down torque on the star, relative to
the classical assumption of a closed field.  Thus, the equilibrium
spin rate predicted by the disk locking picture is much faster, which
calls into question whether disk locking can explain the existence of
slowly rotating accretors.

However, since the MP05 model could only calculate the torque for a
given set of parameters (at a given epoch), their conclusion is based
on the calculation of the equilibrium spin rate.  MP05 were not able
to describe how the spin evolves in time.  Therefore, there is a
question as to whether the the stars actually evolve near equilibrium,
and whether the evolution of the system, including changes in the
accretion rate, stellar radius, and stellar spin rate may affect this
conclusion.  Also, it is important to develop a physical model for
understanding the evolution of stellar spin over a range of ages.  For
these reasons, in this paper, we develop and utilize a stellar spin
evolution model \citep[similar to CC93; AC96;][]{yi94, yi95} that
includes the updated torque theory of MP05.  In this first effort, we
do not attempt to explain all phenomena related to young star spins.
Rather, we focus on testing whether models can produce spin rates
within the observed typical range of $\sim1$--10 days.  This work
extends the MP05 formulation to the time domain, during the first
$3\times 10^{6}$ yr (i.e., during the Hayashi phase).

Section \ref{sec_model} contains a description of the model and
details of our calculations.  Section \ref{sec_results} contains the
results of stellar spin calculations that include the effect of
magnetic field opening.  A discussion and summary of the conclusions
of this work are contained in section \ref{sec_conclusion}.  As a test
of the model, and for illustrative purposes, we include an Appendix
containing the results of stellar spin calculations under the
classical assumption of a completely closed field.

\section{Stellar Spin Evolution Model} \label{sec_model}

The goal of the present work is to determine the effect of different
star-disk interaction torques on the evolution of the spin rates of
young accreting stars.  For this purpose, in addition to calculating
the torques, it is necessary to follow the evolution of the stellar
mass ($M_*$), radius ($R_*$), and moment of inertia, $I_* \equiv k^2
M_* R_*^2$, where $k^2$ is the normalized radius of gyration.  This
section describes the complete model, assumptions, and adopted
parameters.

All of our models follow a one solar mass protostar as it evolves
along the Hayashi (fully convective) track, before the formation of a
radiative core.  The ``birthline'' age of a one solar mass star is
$\sim10^5$~yr \citep{stahler83}, which corresponds to the youngest
ages at which stars are observed.  To allow the system some time to
respond to the initial conditions, we begin our computation at a
somewhat earlier time of $3\times10^4$~yr.  We follow the evolution
until an age of three million years, which is approximately the end of
the Hayashi track phase \citep[e.g.,][]{siess3ea00}.

     \subsection{Mass Accretion} \label{sec_mdot}

The torques acting on the star (detailed in \S \ref{sec_sdint}) arise
from the interaction between the star and a surrounding accretion
disk.  As far as this interaction is concerned, one of the most
important properties of the surrounding disk is the accretion rate.
Specifically, this is the rate at which matter is fed into the
interaction region (within a distance of several $R_*$), which we also
assume is the rate at which material accretes onto the stellar
surface. To follow the decrease of the accretion rate in time, we
adopt an exponential decay \citep[also used by CC93;][]{yi94, yi95},
\begin{equation}
\label{eq_mdot}
\dot{M_{a}} = \frac{M_D}{t_a} \;\; e^{-t/t_a},
\end{equation}
where $t_a$ is the decay timescale and $M_D$ is the ``disk mass,''
equal to the total amount of mass that would accrete from time $t = 0
\rightarrow \infty$.  We adopt $t_a = 10^6$ yr, for all models.

\begin{figure}
\epsscale{\psiz}
\plotone{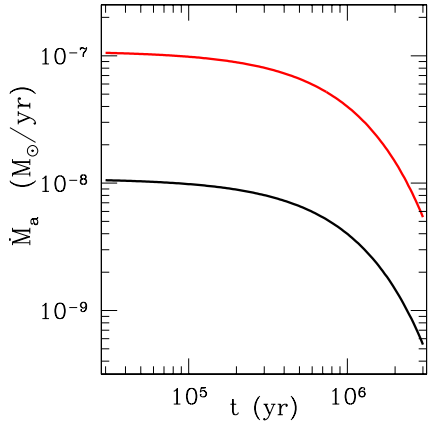}
\caption{Two different accretion histories considered in the models,
  following equation (\ref{eq_mdot}).  Here and in all figures, a red
  line corresponds to the case with the highest accretion rate
  considered, while a black line is for the lowest accretion rate.}
\label{fig_mdot}
\end{figure}

Figure \ref{fig_mdot} shows the evolution of the accretion rate, given
by equation (\ref{eq_mdot}).  In all of our models, we consider two
different disk masses, $M_D = 0.1$ and 0.01 $M_\odot$, to sample a
range of accretion rates.  We chose these two disk masses to represent
a relatively high and low accretion rate, shown as a red and black
solid line in figure \ref{fig_mdot}, respectively.  The true accretion
histories of young stars are likely to be much more complex than
equation (\ref{eq_mdot}), but this is not yet well-understood.  The
advantage of our approach is that it is simple, and it samples the
wide range of accretion rates that is observed in optically visible
accreting young stars \citep[e.g.,][]{hartmannea98,
  siciliaaguilarea05, natta3ea06}.

\begin{figure}
\epsscale{\psiz}
\plotone{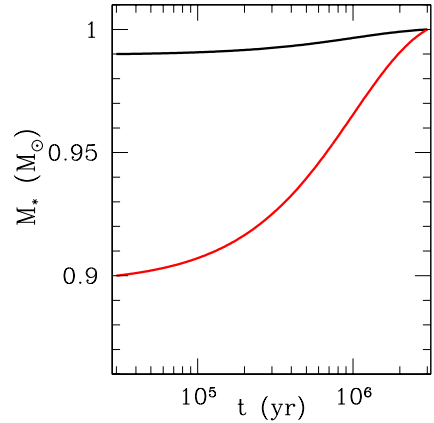}
\caption{Evolution of stellar mass, normalized to solar units, for the
  two different accretion histories considered.  The red and black
  lines represent the high and low accretion rates, respectively.  For
  each case, we chose the initial stellar mass such that the final
  mass would equal one solar mass.}
\label{fig_mstar}
\end{figure}

Figure \ref{fig_mstar} shows the evolution of the stellar mass, which
only depends on the prescribed accretion rate (eq.\ [\ref{eq_mdot}])
and an initial condition (at $t_0=3\times10^4$ yr).  For each of the two
different accretion rates considered here, we set the initial stellar
mass such that the final mass equals one solar mass.  That is, the
initial stellar mass equals one solar mass minus $M_D$, corresponding
to 0.9 and 0.99 $M_\odot$ for the two cases.

     \subsection{Stellar Structure and Evolution} \label{sec_rstar}

For the evolution of stellar radius, we adopt the simple treatment of
CC93 and \citet{yi94, yi95}.  This treatment models the structure of
the star as a polytrope with an index of $n = 3/2$ and assumes a
constant photospheric temperature, $T_e$, during the Hayashi phase.
The polytropic model results in a mean radius of gyration of $k^2 =
0.2$.  The radiated, blackbody luminosity of the star is powered only
by the release of gravitational potential energy, as the star
contracts and accretes matter.  The evolution of the stellar radius is
then given by
\begin{equation}
\label{eq_rstar}
\frac{dR_{*}}{dt} = 2\frac{R_{*}}{M_{*}}\dot{M_{a}} -
\frac{28\pi\sigma R_{*}^4 T_{e}^4}{3GM_{*}^2},
\end{equation}
where $G$ is Newton's grativational constant and $\sigma$ is the
Stefan-Boltzmann constant.

\begin{figure}
\epsscale{\psiz}
\plotone{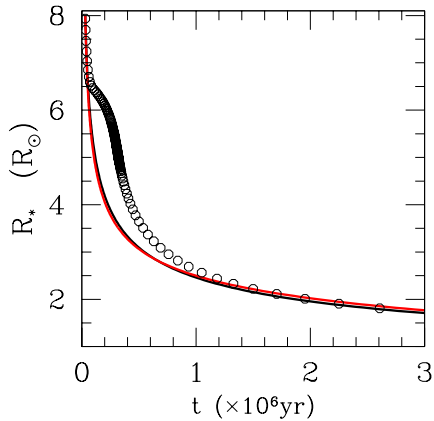}
\caption{Evolution of stellar radius normalized to solar units.  The
  solid lines show solutions of equation (\ref{eq_rstar}), with the
  black and red lines corresponding to the low and high accretion
  rates (respectively) shown in figure \ref{fig_mdot} and discussed in
  section \ref{sec_mdot}.  For comparison, the circular symbols show
  the prediction of a more detailed model by \citet{siess3ea00}.}
\label{fig_rstar}
\end{figure}

The black and red solid lines in figure \ref{fig_rstar} show the
evolution of stellar radius with time, resulting from equations
(\ref{eq_mdot}) and (\ref{eq_rstar}).  In all of our models, we
adopted an initial stellar radius of $8R_\odot$ and a photospheric
temperature of $T_e=4280$ K.  We chose these values in order to best
match the radius evolution of a one solar mass star (with Z=0.02)
predicted by the more sophisticated stellar model of
\citet[][]{siess3ea00}.  Also, the chosen value of $T_e$ is within the
narrow range of (nearly constant) Hayashi track temperatures exhibited
by the \citet{siess3ea00} model.  For comparison, the circles in the
figure show the \citet{siess3ea00} model predictions.

It is clear from figure \ref{fig_rstar} that there is a ``bump'' in
the Siess et al.\ radius evolution from $\sim1$--$5\times10^5$~yr that
is not matched by our model.  This is due to deuterium burning, which
we neglect.  Still, our simple treatment matches the Siess et
al.\ model at the begining and end of the evolution, demonstrating
that the overall evolution of stellar radius during this phase is
dominated by the Kelvin-Helmholz contraction.  Thus, our simple
treatment is sufficient for the present purpose.  Also, the treatment
of equation (\ref{eq_rstar}) captures the effects of mass accretion,
which is neglected in more sophisticated models \citep[with the
  exception of][]{tout3ea99}.  To compute the evolution for a longer
time, beyond the Hayashi track, would require the use of a more
sophisticated stellar model, which we leave for future work.

     \subsection{Spin Evolution} \label{sec_spin}

Along with the prescription above for the evolution of the star's
mass, radius, and moment of inertia, the model also computes the
angular momentum evolution.  Assuming the star rotates as a solid
body, the angular momentum equation can be expressed in terms of spin
evolution,
\begin{eqnarray}
\label{eq_angmom}
\frac{d\Omega_{*}}{dt} = \frac {T_{*}}{I_*} - 
\Omega_{*} \left( \frac{\dot M_a}{M_*} + \frac{2}{R_*} \frac{dR_*}{dt} \right), 
%
%
\end{eqnarray}
where $\Omega_{*}$ is the angular velocity of the star, and $T_{*}$ is
the net torque on the star, discussed in section \ref{sec_sdint}.

It is sometimes convenient to express the spin rate as a fraction of
the breakup speed, defined as the Keplerian velocity at the star's
equator.  This normalized spin rate is defined
\begin{eqnarray}
\label{eq_f}
f \equiv \Omega_* \sqrt{\frac{R_*^3}{G M_*}}.
\end{eqnarray}
In all of our models, we will consider two cases with different
initial spin rates.  The two cases have initial fractional spin rates
of $f_0 = 0.3$ and 0.06, representing the two extremes of rapid and
slow initial rotation.  In all of our models, we enforce a maximum
spin rate of $f=1$, since at this spin rate, some of the model
assumptions (e.g., regarding the stellar structure) break down.  As we
show below, this limit is only reached in some of the most extreme
models.

     \subsection{Star-Disk Interaction Torques} \label{sec_sdint}

To calculate the torque on the star arising from accretion and the
magnetic interaction between the star and disk, we follow the
formulation of MP05.  Here, we summarize the parts of that work that
are relevant for our spin evolution model, and the reader will find
details in that paper.  This torque model follows several previous
works \citep[e.g., CC93; AC96;][]{ghoshlamb78, yi94, yi95,
  lovelace3ea95}, but the primary advantage the MP05 formulation has
for this work is that it includes the effects of varying magnetic
coupling and magnectic connectedness between the star and disk.

          \subsubsection{Magnetic Field Prescription} \label{sub_magfield}

The basic model is one-dimensional (in radius, $r$) and assumes the
star has a rotation-axis-aligned dipolar magnetic field anchored into
its surface, such that the magnetic field strength varies in the
equatorial plane as
\begin{eqnarray}
\label{eq_bdip}
B_z = B_* \left(\frac{R_*}{r}\right)^3,
\end{eqnarray}
where $B_*$ is the magnetic field strength at the equator and surface
of the star.  The surrounding accretion disk is assumed to be thin, so
that the magnetic field has a negligible radial component along the
disk surface.

The torque depends strongly upon the strength of the large-scale
magnetic field $B_*$.  Both disk locking and stellar wind spin-down
models for T Tauri stars require global (dipolar) magnetic fields with
surface strengths in the range of hundreds to thousands of Gauss.
Observations of pre-main-sequence stars typically find field strenghts
with a mean absolute value of $\sim1$--3~kG
\citep[e.g.,][]{basri3ea92, johnskrull07, johnskrullea09}, while the
large-scale (global) fields of these stars appear to be at most
several hundred Gauss \citep{safier98, johnskrullea99, smirnovea04,
  yang3ea07, bouvierea07, donatiea07, donatiea08, hussainea09}.  While
this discrepancy means that the magnetic fields are complex (not
simple dipoles), it is clear that these stars possess dynamically
significant magnetic fields \citep{gregoryea08}.  For the models
presented here, we choose two values of the magnetic field strength,
$B_*$ = 500 G and 2000 G, in addition to some cases with $B_*=0$, used
for comparison.

Previous models for pre-main-sequence spin evolution \citep[e.g.,
  CC93; AC96;][]{yi94, yi95} chose a prescription for the magnetic
field that depends upon the stellar spin rate and stellar radius, as
motivated by the expectations of a stellar dynamo.  By contrast, for
the models presented here, we choose to keep the magnetic field
constant in time (a constant field was also considered in the works of
\citealp{johnskrullgafford02} and \citealp{yi94}) for three reasons.
First, observations of mean fields of T Tauri stars find a relatively
constant field strength \citep{johnskrullgafford02, johnskrull07},
regardless of stellar age, radius, or spin rate.  Measurements of
large-scale field \citep{johnskrullea99, smirnovea04, yang3ea07,
  bouvierea07, donatiea07, donatiea08, hussainea09} are more rare, and
there are not yet enough clear detections to look for trends.  So a
fixed field is consistent with observations.  Second, the observed
relationship between X-ray activity and spin rates of T Tauri stars
\citep{stassunea04} suggests that they are primarily in the
``saturated'' or ``supersaturated'' regime \citep{johnskrull07}, where
solar-like stellar dynamo relationships are thought to break down.  It
is not yet clear how the global magnetic field behaves in this fully
convective phase \citep{browning08, donatiea08}, but at this point,
adopting a constant field strength seems appropriate.  Finally, the
main purpose of the present work is to demonstrate how the stellar
spin rate is influenced by the magnetic coupling to the disk.  Thus,
it is also convenient for illustrative purposes to keep the field a
fixed constant.


           \subsubsection{Magnetic Coupling to the Disk} \label{sub_coupling}

The stellar magnetic field lines vertically threading the disk are
imperfectly coupled to it.  To be clear, throughout this paper, the
term ``coupling'' refers to the extent to which the magnetic field is
``frozen into'' the disk material---that is, the effective magnetic
diffusion in the disk.  By contrast, the magnetic ``connection''
refers to the magnetic field loops that are attached both to the star
and disk.  MP05 showed how the strength of the coupling affects the
amount of connected flux, and how this in turn affects the integrated
torques.

The disk material rotates with Keplerian speed, which means that there
is a singular location where the angular rotation rate of the disk
equals that of the star, the corotation radius,
\begin{eqnarray}
\label{eq_rco}
R_{\rm co} \equiv \left(\frac{G M_*}{\Omega_*^{2}}\right)^{1/3} = f^{-2/3} R_*.
\end{eqnarray}
Every location, except for $R_{\rm co}$, rotates at a different
angular rate than the star.  Thus, the magnetic field connected to the
star and threading the disk will be twisted azimuthally (in the
$\phi$-direction).  It is assumed that the material above both the
disk and star (the ``corona'') is of sufficiently low density (high
Alfv\'en speed) that the twisting of the magnetic field at the disk
surface is equilibrated along the field line on a rapid timescale.
Thus, the twisting of the magnetic field by the differential rotation
between the star and disk imparts a torque, transferring angular
momentum between the two.

The variable $\gamma(r) \equiv B_\phi/B_z$ quantifies the twist of the
field at the surface of the disk at each radial location.  This
twisting occurs rapidly, on a stellar rotation timescale, so that a
steady-state $\gamma(r)$ is simply determined by the balance between
the differential rotation and the tendancy for the magnetic field to
untwist by slipping (or reconnecting) azimuthally through the disk.
This slipping rate depends on how well the magnetic field is coupled
to the disk material.

To describe the coupling, MP05 adopted a dimensionless magnetic
diffusivity parameter, which they assume to be constant throughout the
disk,
\begin{eqnarray}
\label{eq_beta}
\beta \equiv \frac{\eta_t}{h v_k},
\end{eqnarray}
where $\eta_t$ is the effective magnetic diffusivity, $h$ is the
vertical thickness of the disk, and $v_k$ is the Keplerian rotation
velocity.  In other words, $\beta$ is the inverse of the effective
magnetic Reynolds number of the star-disk interaction.  A large value
of $\beta$ corresponds to weak magnetic coupling (rapid field slipping
through the disk), and a small value to strong coupling (slow
slipping).  The various models of star-disk interaction torques in the
literature differ in their treatment of the magnetic field coupling,
but all models adopt parameters that give similar results to MP05 with
a value of $\beta$ of order unity\footnote{The one exception to this
  is the model of CC93 (see the Appendix).}.  However, $\beta$ is a
highly unknown parameter, and MP05 suggested that $\beta \la 10^{-2}$
(i.e., large magnetic Reynolds number) may be more reasonable for
accreting pre-main-sequence stars.  In our models presented here, we
consider $\beta=10^{-2}$ to represent the most realistic case, but we
also show cases with $\beta=1$, for illustrative purposes.

          \subsubsection{Magnetic Connection State and $\gamma_c$} 

\label{sec_state}

There is a dynamical limit to how strongly a dipolar magnetic field
can be twisted azimuthally.  As shown by several authors
\citep[][]{Aly85, Alykuijpers90, lyndenbellboily94, lovelace3ea95,
  bardouheyvaerts96, agapitoupapaloizou00, uzdensky3ea02}, when the
absolute value of the twist $\gamma$ exceeds a value near unity, the
magnetic pressure force associated with $B_\phi$ pushes outward and
leads to an opening of the dipole field loops.  For practical
purposes, this simply means that the star and disk are no longer
causally linked, and no torques can be transmitted between the two.
This does not mean the open field lines necessarily impart zero
torque.  Open field lines can, and likely do, transport some angular
momentum from the star and/or disk (via winds).  However, no angular
momentum is exchanged between the star and disk along open field
lines.  In the present work, we consider only the torques arising in
the star-disk interaction and thus neglect any torques that may arise
from open field lines.

To take into account the opening of the field, the MP05 formulation
includes a maximum for the absolute value of the twist, $\gamma_c$.
In the regions of the disk where the twist is greater than this
critical value, the model assumes that the magnetic field no longer
connects the star to the disk.  The torque on the star that would have
arisen from those magnetic field lines is instead taken to be zero.
MP05 showed that a value of $\gamma_c = \infty$ represents the
classical assumption of a field that remains connected at all radii in
the disk.  To show the effect of the opening of the magnetic field in
more realistic systems, a value of $\gamma_c = 1$ is appropriate
\citep{uzdensky3ea02}.

Furthermore, MP05 showed that there exists a mode change in the
magnetic connection between the star and disk, at a threshold value of
the spin rate.  Specifically, if
\begin{eqnarray}
\label{eq_fcrit}
f < (1-\beta \gamma_c) (\gamma_c \psi)^{-3/7},
\end{eqnarray}
where
\begin{eqnarray}
\label{eq_psi}
\psi \equiv \frac{2 B_*^2 R_*^{5/2}}{\dot M_a \sqrt{G M_*}},
\end{eqnarray}
then the stellar magnetic field only connects to a small region near
the inner edge of the disk.  In this case, which they call ``State
1,'' no magnetic field connects the star to a radius in the disk that
is larger than $R_{\rm co}$.  This is significant, since in this state
there exist no star-disk interaction torques that act to spin the star
down (i.e., there are only spin-up torques).  Alternatively, when
equation (\ref{eq_fcrit}) is not valid (for faster spin), the system
is in ``State 2,'' characterized by a magnetic connection to the disk
over a range of radii on either side of $R_{\rm co}$.  In State 2, the
star experiences both spin-up and spin-down torques.  Figure 3 of MP05
illustrates these two magnetic connection states \footnote{In the
  present work, we do not consider ``State 3'' discussed by MP05,
  which represents the ``propeller'' regime
  \citep{illarionovsunyaev75} in which there is no accretion of
  material onto the star.}.

As evident in equation (\ref{eq_fcrit}), magnetic connection State 1
only occurs for relatively extreme cases of slow spin rate, weak
field, and/or high accretion rate.  As it turns out, all of the models
with non-zero $B_*$ presented in this paper remain in State 2 for the
entire time simulated.

          \subsubsection{Truncation Radius and Accretion Torque} \label{sec_rt}

In the region of the disk at smaller radii than $R_{\rm co}$, the
twist of the field is such that torques remove angular momentum from
the disk (giving it to the star).  If the magnetic field is strong
enough, there will be a radius above the stellar surface at which
these torques can extract all of the angular momentum of the material
in the disk at that radius.  Under these circumstances, the disk will
be truncated, and accretion will proceed as a free-fall of material
onto the star along the dipolar magnetic field lines.

In State 1 [eq.\ (\ref{eq_fcrit})], the location of the truncation
radius is determined by
\begin{eqnarray}
\label{eq_rt1}
{\rm (State\; 1)} \;\;\;\;\;\;\;\;\;\;\;\;\;
R_t = (\gamma_c \psi)^{2/7} R_*.
%
\end{eqnarray}
In State 2, the location of the truncation radius obeys
\begin{eqnarray}
\label{eq_rt2}
{\rm (State\; 2)} \;\;\;\;
\left(\frac{R_t}{R_{\rm co}}\right)^{-7/2}
\left[1-\left(\frac{R_t}{R_{\rm co}}\right)^{3/2}\right] =
\frac{\beta}{\psi f^{7/3}},
\end{eqnarray}
which we solve for $R_t < R_{\rm co}$ using a Newton-Raphson method.
In our models, for each timestep, we first use equation
(\ref{eq_fcrit}) to determine the magnetic connection state, then
either equation (\ref{eq_rt1}) or (\ref{eq_rt2}) to find $R_t$.
Finally, we check that $R_t$ is greater than $R_*$.  If not, this
means the magnetic field is not strong enough to truncate the disk
above the stellar surface, and we reset $R_t$ to equal $R_*$.

The truncation of the disk is due to the stellar magnetic field
extracting the angular momentum of disk material at a radius of $R_t$.
Thus, the truncation of the disk and subsequent accretion of material
adds angular momentum to the star at a rate equal to
\begin{eqnarray}
\label{eq_ta}
T_a = \dot M_a \sqrt{G M_* R_t},
\end{eqnarray}
which is referred to as the ``accretion torque''\footnote{Equation
  (19) of MP05 for the accretion torque had an extra (negligible) term
  in brackets of $k^2f$, which is not correct.  This term actually
  represents a term already in the basic angular momentum equation,
  which is due to the change in the stellar moment of inertia
  (specifically, the $\Omega_* k^2 R_*^2 \dot M_a$ term).  Equation
  (\ref{eq_ta}) of this work, which has also been derived by many
  previous authors \citep[e.g., CC93; AC96;][]{ghoshlamb79b},
  represents the physical accretion torque.}.  Note that, although
equation (\ref{eq_ta}) does not explicitly contain a dependence on the
magnetic field strength, the magnetic field is central to the
determination of $R_t$ (as long as $R_t > R_*$).  Furthermore, when
$R_t > R_*$, material falling onto the star transfers the bulk of its
angular momentum to the magnetic field before the material reaches the
stellar surface \citep[e.g.,][]{romanovaea02, long3ea05}.  Thus, as
long as $R_t > R_*$, the torque of equation (\ref{eq_ta}) is
experienced by the star as a magnetic torque, although we still refer
to this as the ``accretion torque'' to distinguish it from the
component of the torque described in section \ref{sec_tmag}.

          \subsubsection{Magnetic Torque} \label{sec_tmag}

In addition to the accretion torque, if the system is in State 2, the
magnetic connection over a range of radii in the disk transports
angular momentum between the star and the disk.  In this case
\begin{eqnarray}
\label{eq_tmag2}
{\rm (State\; 2)} \;\;\;\;
T_m = \frac{B_*^2 R_*^6}{3 \beta R_{\rm co}^3}
[2(1+\beta\gamma_c)^{-1} - (1+\beta\gamma_c)^{-2}
\nonumber \\
- 2(R_{\rm co}/R_t)^{3/2} + (R_{\rm co}/R_t)^3],
\;\;\;\;\;
\end{eqnarray}
is the net torque felt by the star from both spin-up and spin-down
components.  We refer to this as the ``magnetic torque.''  If the
system is in State 1, the magnetic field does not connect to a
significant range of radii in the disk, so we set
\begin{eqnarray}
\label{eq_tmag1}
{\rm (State\; 1)} \;\;\;\;\;\;\;\; 
T_m = 0.
\end{eqnarray}

As discussed in MP05, we assume that the disk will be capable of
transporting away any angular momentum that it receives from the star
(i.e., from negative torques in eq.\ [\ref{eq_tmag2}]).  For example,
the disk may restructure itself in response to external torques
\citep[e.g.,][]{rappaport3ea04} in such a way as to increase the
effectiveness of viscous or turbulent angular momentum transport.



     \subsubsection{Total Torque and Equilibrium Spin Rate} \label{sub_equilib}

\begin{figure*}
\epsscale{\psizf}
\plottwo{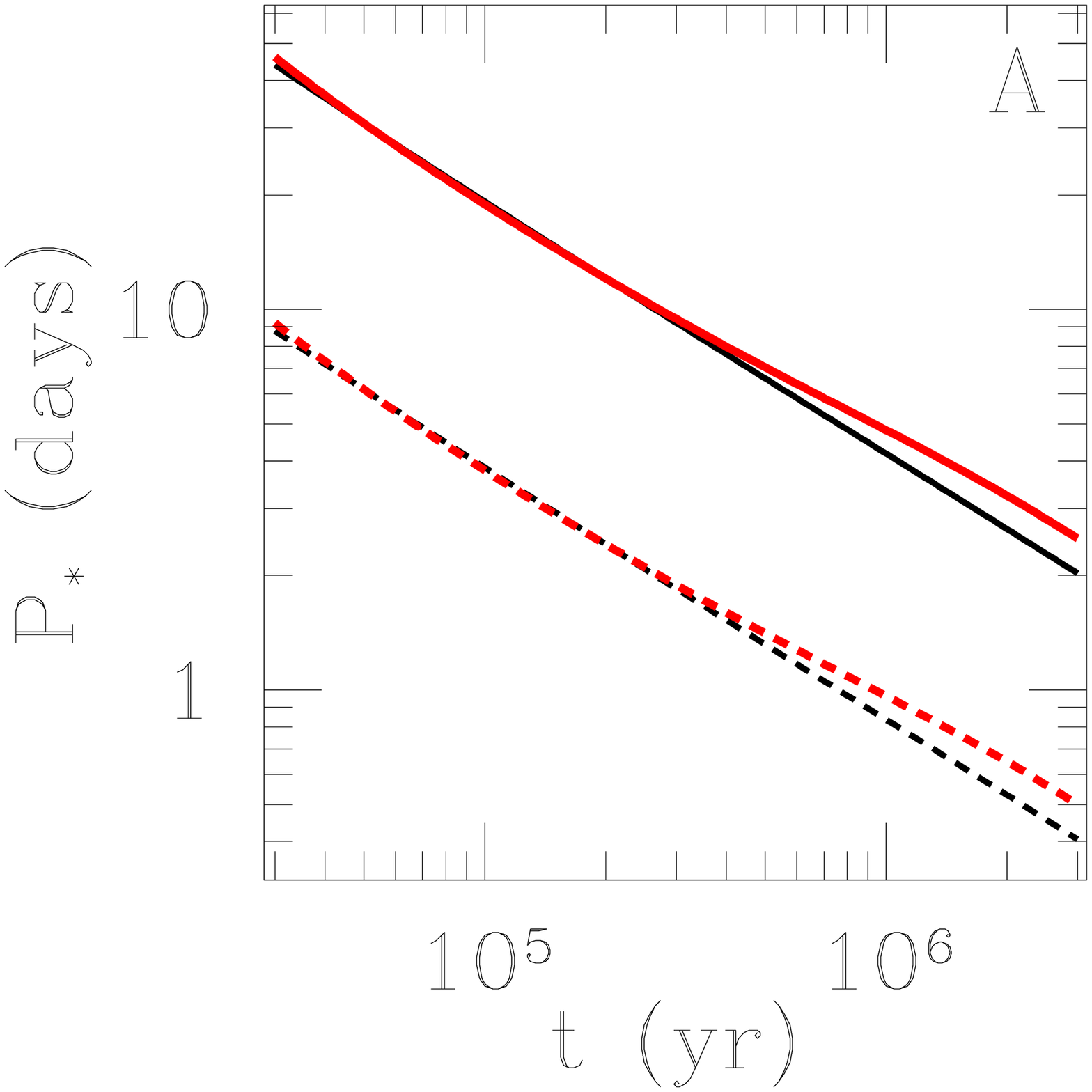}{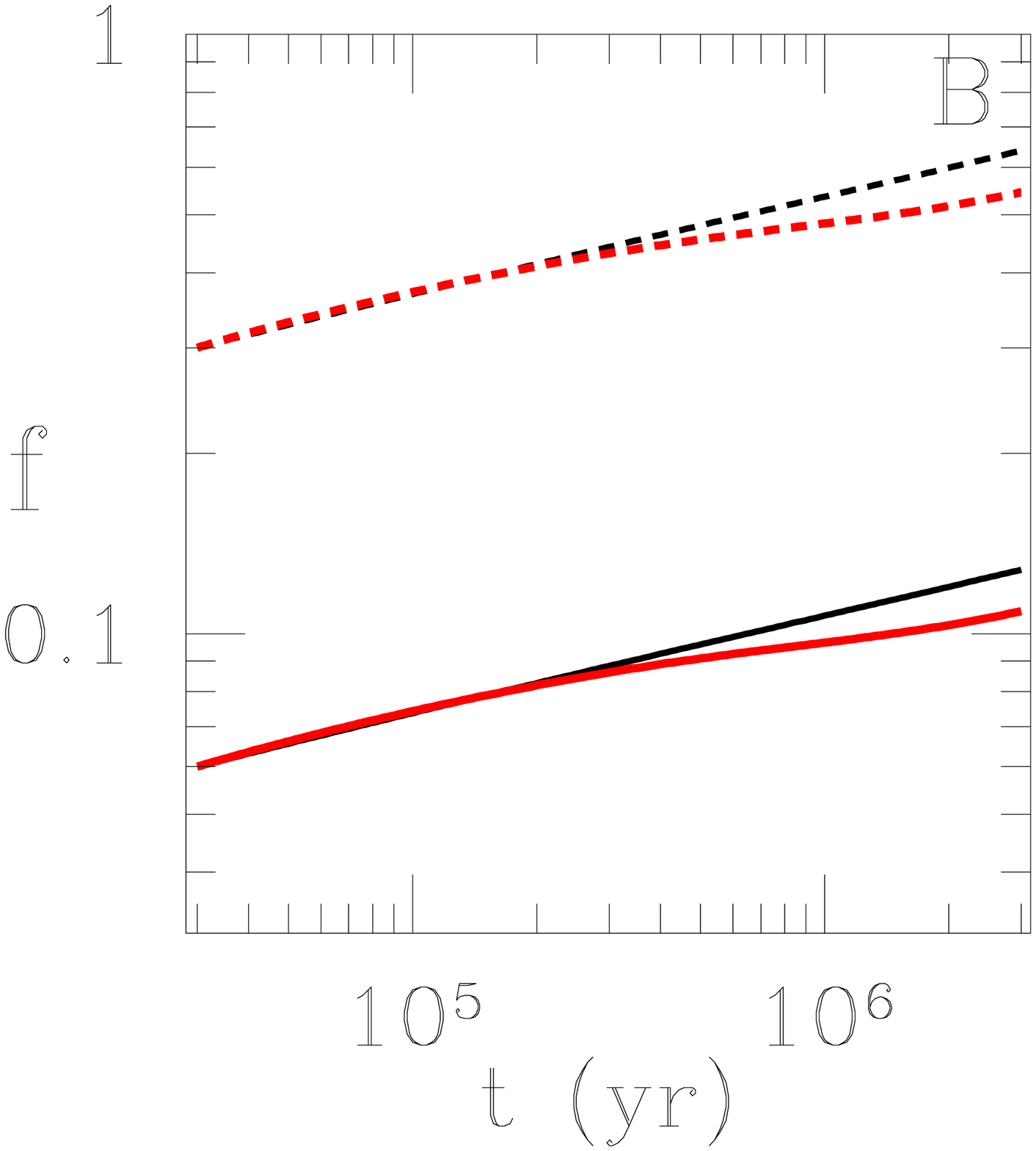}
\caption{Evolution of spin period (panel a) and spin rate expressed as
  a fraction of the breakup rate (panel b) for a star with no external
  torques ($T_*=0$).  The four cases shown represent the two choices
  of accretion history and the two initial spin rates.  The red and
  black lines correspond to the high and low accretion cases,
  respectively.  The solid and dashed lines correspond to models with
  an initial spin rate of 0.06 and 0.3 times the breakup rate,
  respectively.}
\label{fig_t0}
\end{figure*}

The total torque on the star is the sum of the accretion and magnetic
torques,
\begin{eqnarray}
\label{eq_torque}
T_* = T_a + T_m,
\end{eqnarray}
and we neglect any other torques that may be present (e.g., from a
stellar wind).  The total torque can be either postive or negative,
depending on the balance between spin-up torques (due to the trucation
of the disk and to magnetic field lines that are azimuthally twisted
in a direction that leads the rotation of the star) and spin-down
torques (due to magnetic field lines that are twisted in a direction
that lags the rotation of the star).

For any given system, there is a particular stellar spin rate at which
the net torque will equal zero ($T_*=0$).  This is called the
equilibrium spin rate, and the torque always acts to change the
stellar spin in a direction toward this equilibrium.  As indicated in
equation (\ref{eq_angmom}), the spin also changes in response to
changes in stellar radius and mass, and in general to changes in the
moment of inertia.  If the contribution from the torque is large
compared to that from changes in the moment of inertia, the stellar
spin will evolve asymptotically toward the equilibrium rate.

This general idea provides the theoretical underpinning for the idea
of disk locking \citep{ghoshlamb78, camenzind90, konigl91,
  edwardsea93, shuea94}, which holds that the stellar spin rate is
determined by the current conditions in the star-disk interaction and
independent of any initial conditions.  MP05 described the method of
computing the stellar spin rate in this hypothetical equilibrium state
for their torque formulation, which uses equations (23)--(27) of that
work.  In presenting our results below, we compare the actual spin
rates to the equilibrium values to gain insight into these systems.

    \subsection{Numerical Method}

The coupled equations (\ref{eq_mdot}), (\ref{eq_rstar}), and
(\ref{eq_angmom}) describe the evolution of the system. We wrote a
computational code that solves these simultaneously, using the
fourth-order Runge-Kutta scheme of \citet{pressea94}, starting from
$t_0=3\times10^4$ yr and ending at 3 Myr.  For computational
efficiency and numerical stability, we use a dynamic timestep.  At
each step, we compute the next timestep by requiring that none of the
three main variables, $\Omega_{*}$, $R_{*}$, $M_{*}$, change by more
than 1\% per step.  This gives results converged to three significant
figures and typically requires a few hundred timesteps for the entire
evolution.

In order to compute the torque at each timestep, the code first solves
equation (\ref{eq_fcrit}) to determine the magnetic connection state
of the system.  Then, depending on the state, the code solves either
equation (\ref{eq_rt1}) or (\ref{eq_rt2}) to determine the disk
truncation radius.  We enforce a minimum value of $R_t=R_*$.  Finally,
the code calculates the accretion torque using equation (\ref{eq_ta})
and the magnetic torque using either equation (\ref{eq_tmag2}) or
(\ref{eq_tmag1}).  The torque is assumed to be constant during each
timestep.  We also enforce a maximum on the spin rate, corresponding
to $f=1$ (see \S \ref{sec_spin}).

\section{Results} \label{sec_results}

\begin{deluxetable}{ccccccc}
\tablewidth{0pt}
\tablecaption{Model Parameters \label{tab_parms}}
\tablehead{
\colhead{Case} &
\colhead{$\gamma_c$} &
\colhead{$\beta$} &
\colhead{$B_*$ (Gauss)} &
\colhead{$M_{D}/M_{\odot}$} &
\colhead{$f_0$} &
\colhead{Figure} 
}

\startdata

$T_*=0$ & \nodata & \nodata & \nodata & 0.01, 0.1 & 0.06, 0.3 & \ref{fig_t0} \\
$B_*=0$ & \nodata & \nodata & 0       & 0.01, 0.1 & 0.06, 0.3 & \ref{fig_b0} \\
O1   & 1       & 0.01    & 500      & 0.01, 0.1 & 0.06, 0.3 & \ref{fig_g1b500bet01} \\
O2   & 1       & 0.01    & 2000     & 0.01, 0.1 & 0.06, 0.3 & \ref{fig_g1b2000bet01} \\
O3   & 1       & 1       & 2000     & 0.01, 0.1 & 0.06, 0.3 & \ref{fig_g1b2000bet1} \\
C1 & $\infty$& 1       & 500     & 0.01, 0.1 & 0.06, 0.3 & \ref{fig_ginfb500bet1}  \\
C2 & $\infty$& 1       & 2000    & 0.01, 0.1 & 0.06, 0.3 & \ref{fig_ginfb2000bet1} 

\enddata

\end{deluxetable}

This section contains results from several models.  Table
\ref{tab_parms} lists the parameters for all cases, in order of their
presentation and grouped by the figures in which the results appear.
All models have the same initial stellar radius ($8R_\odot$) and
effective temperature ($T_e=4280$~K; see fig.\ \ref{fig_rstar} and \S
\ref{sec_rstar}).  Also, all models end with a mass of $M_\odot$, so
that the initial mass depends on the accretion rate (see \S
\ref{sec_mdot}).  For each case in table \ref{tab_parms}, we ran four
models, one for each combination of two different initial spin rates
(see \S \ref{sec_spin}) and two different mass accretion rates (see \S
\ref{sec_mdot}).  These parameters are intended to represent a range
that is appropriate for accreting T Tauri stars.

Section \ref{sec_t0} includes two simplified cases with no magnetic
fields: one in which $T_*=0$, so that there is no star-disk
interaction; and one in which $B_*=0$ so that there are no magnetic
effects, just disk accretion.  Section \ref{sec_gam1} contains the
main results, models that include magnetic fields with realistic field
line opening ($\gamma_c=1$) and different values of $\beta$, for two
different field strengths.  For comparison, and as a verification of
our model and computations, the appendix contains results for cases
with the classical assumption of a fully closed magnetic field
($\gamma_c=\infty$) and $\beta=1$.

     \subsection{$T_*=0$ and $B_*=0$ Cases} \label{sec_t0}

\begin{figure*}
\epsscale{\psiz}
\plotone{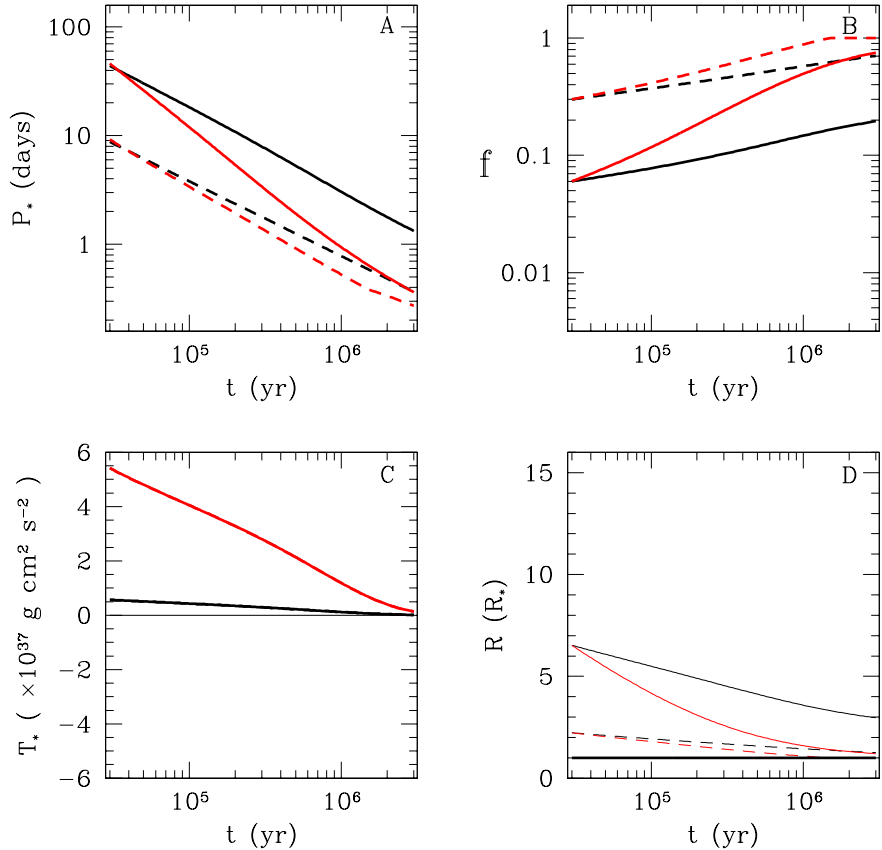}
\caption{Results for $B_*=0$.  Panel (a) and (b) show the same
  quantities as figure \ref{fig_t0}.  The solid, dashed, red, and
  black lines in all panels have the same meaning as in figure
  \ref{fig_t0}.  Panel (c) shows the net torque on the star for each
  case.  In this panel, the dashed and solid lines of a given color
  lie on top of each other, since the spin rate does not affect the
  torque when $B_*=0$.  The horizontal dotted line indicates $T_*=0$
  for clarity.  Panel (d) shows the location of the truncation radius
  ($R_t$, thick lines) and the corotation radius ($R_{\rm co}$, thin
  lines; eq.\ [\ref{eq_rco}]), in units of the stellar radius.  Here,
  $R_t=R_*$ for all models, since there is no magnetic field to
  truncate the disk above the stellar surface.}
\label{fig_b0}
\end{figure*}

To begin, it is useful to examine a few simplified cases.  The main
goal is to understand individually how the stellar contraction and
accretion alone affects the spin evolution, before adding the effects
of magnetic fields.  We will consider two cases here: one in which the
net torque on the star is forced to be zero; and one in which there
are no magnetic fields, so that the only torque on the star is due to
the accretion of material from the disk.

Figure \ref{fig_t0} shows the evolution of stellar spin for the case
of zero torque ($T_*=0$ in eq.\ [\ref{eq_angmom}]).  The figure shows
both the spin period, $P_* = 2\pi/\Omega_*$, as well as the spin rate
expressed as a fraction of breakup speed (eq.\ [\ref{eq_f}]), as a
function of time.  In this figure and in each subsequent figure, we
show four different simulations, representing all combinations of two
different accretion rates and two different initial spin rates.  The
red (black) colored lines correspond to the highest (lowest) accretion
rate considered (see \S \ref{sec_mdot} and figure \ref{fig_mdot}).
The solid (dashed) lines correspond to the slow (fast) initial spins
considered (see \S \ref{sec_spin}).

The spin evolution of the stars in figure \ref{fig_t0} is due to
contraction and the accretion of mass containing no angular momentum.
After the initial time of $3\times10^4$ yr, the cases with higher
accretion rates (red lines) spin somewhat more slowly than the cases
with lower accretion rates.  This is because the stars in the high
accretion cases are gaining more mass, while conserving their angular
momentum.  Still, this effect is relatively minor, and it is clear
that the spin evolution is dominated by the change in stellar radius.
Specifically, the contraction from the beginning to the end of the
evolution (see figure \ref{fig_rstar}), leads to a change in spin
period by a factor of approximately 20, and a change in $f$ by a
factor of approximately 2.

Next, we consider the case where angular momentum is added to the star
from the accretion of disk material.  Since the disk is assumed to be
in Keplerian orbit, it has high specific angular momentum relative to
the star.  In the simplified case where the star has no magnetic
field, the disk material will extend all the way to the surface of the
star, forming a boundary layer \citep{lyndenbellpringle74}.  The
torque on the star from the boundary layer is given by equation
(\ref{eq_ta}), with $R_t=R_*$.  This behavior is captured
automatically in our model by setting $B_*=0$.

\begin{figure*}
\epsscale{\psiz}
\plotone{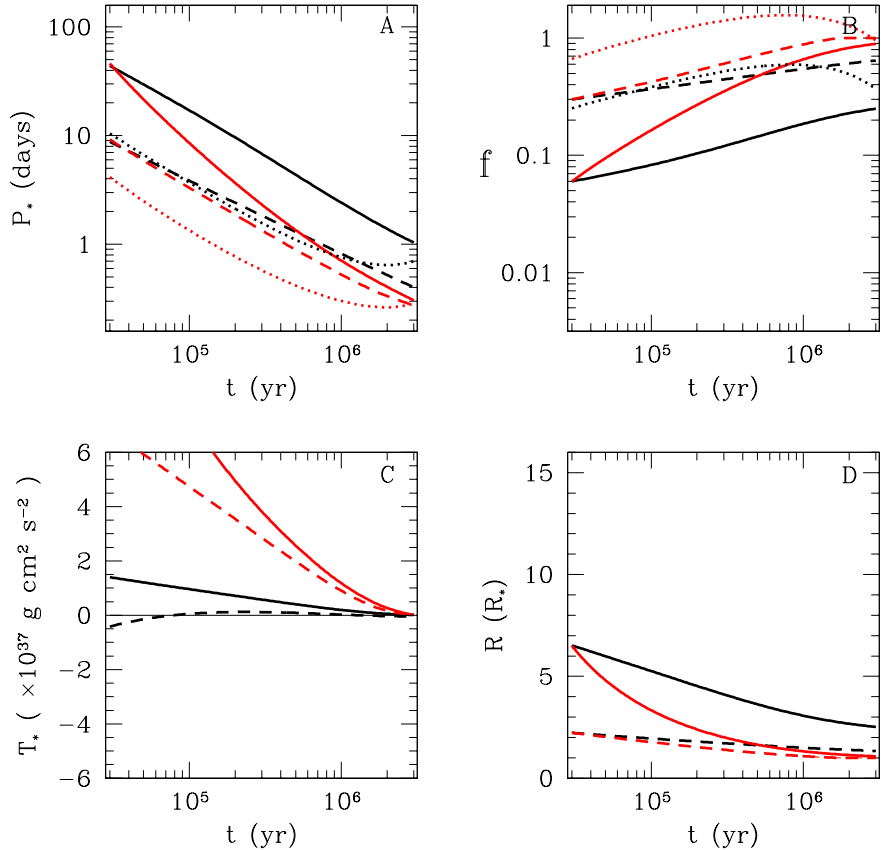}
\caption{Results for $B_*=500$~G, $\gamma_c=1$, and $\beta=0.01$, in
  the same format as figure \ref{fig_b0}.  In panels (a) and (b), the
  dotted red and black lines show the equilibrium value (\S
  \ref{sub_equilib}) for the high and low accretion rates,
  respectively.  In panel (d), the thin and thick lines of a given
  color and type are indistinguishable, since $R_t$ is very close to
  $R_{\rm co}$ in all four cases.}
\label{fig_g1b500bet01}
\end{figure*}

Figure \ref{fig_b0} shows this case for the two different accretion
rates and two different intiial spin rates.  In addition to the
evolution of stellar spin (panels a and b), the figure shows the net
torque on the star (panel c) and the location of the disk truncation
radius and the corotation radius (panel d).  Panel (c) indicates that
the torque on the star is always positive (adding angular momentum to
the star).  The torque does not depend on the spin rate of the star
but depends most strongly on the accretion rate (eq.\ [\ref{eq_ta}]).
Thick lines in panel (d) all overlap, indicating that the disk extends
all the way to the stellar surface ($R_t=R_*$, as expected) in all of
these cases.  The thin lines in panel d show how the corotation radius
(eq.\ [\ref{eq_rco}]) changes with spin rate.

It is clear from figure \ref{fig_b0} how the accretion torque
influences the spin evolution of the star.  For the lowest accretion
rate considered, after 3 Myr, the star spins about 10\% to 50\% faster
than in the $T_*=0$ case (fig.\ \ref{fig_t0}).  But for the highest
accretion rate considered, the stars spin much faster.  The most
extreme case, with a high accretion rate and fast initial spin (dashed
red line), reached the $f=1$ limit enforced by our model (see \S
\ref{sec_spin}) in approximately 1.5 Myr.  In the case with the
slowest initial spin and lowest accretion rate, the spin period is
1--2 days during the last few Myr.  But the other three cases have
spin periods shorter than one day during that time.  In order to
explain the observed spin periods of several days, for the range of
accretion rates considered here, these stars must experience
significant spin down torques.  The next subsection presents cases
that include the additional effects of torques that arise in the
magnetic star-disk interaction.

\begin{figure*}
\epsscale{\psiz}
\plotone{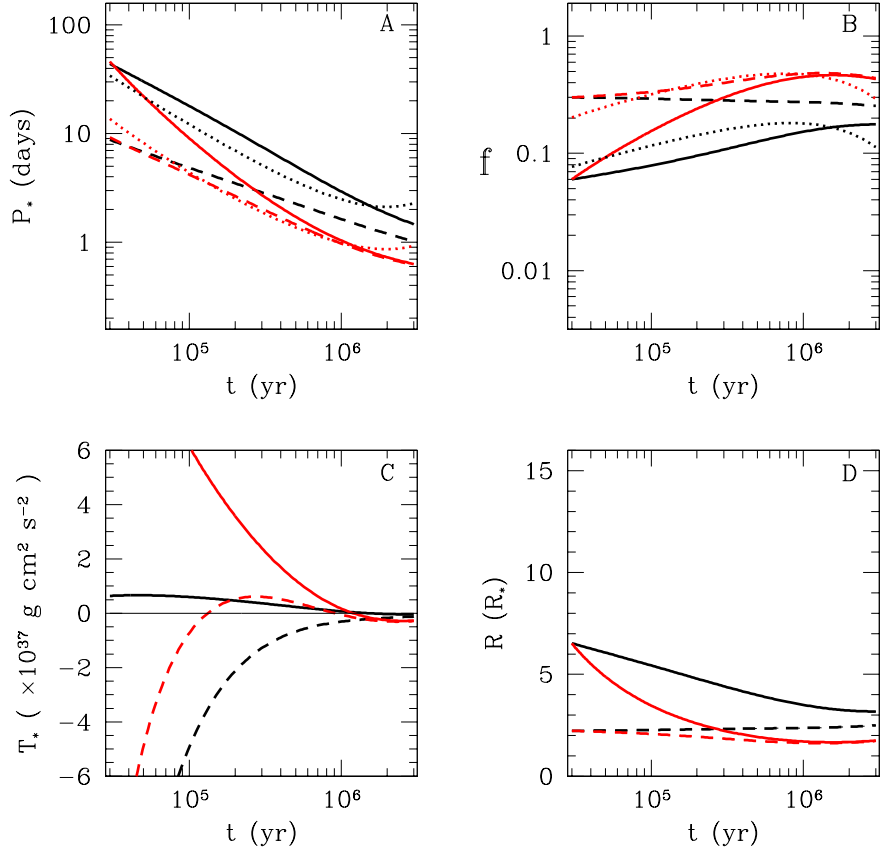}
\caption{Results for $B_*=2000$~G, $\gamma_c=1$, and $\beta=0.01$, in
  the same format as figure \ref{fig_g1b500bet01}.}
\label{fig_g1b2000bet01}
\end{figure*}

\begin{figure*}
\epsscale{\psiz}
\plotone{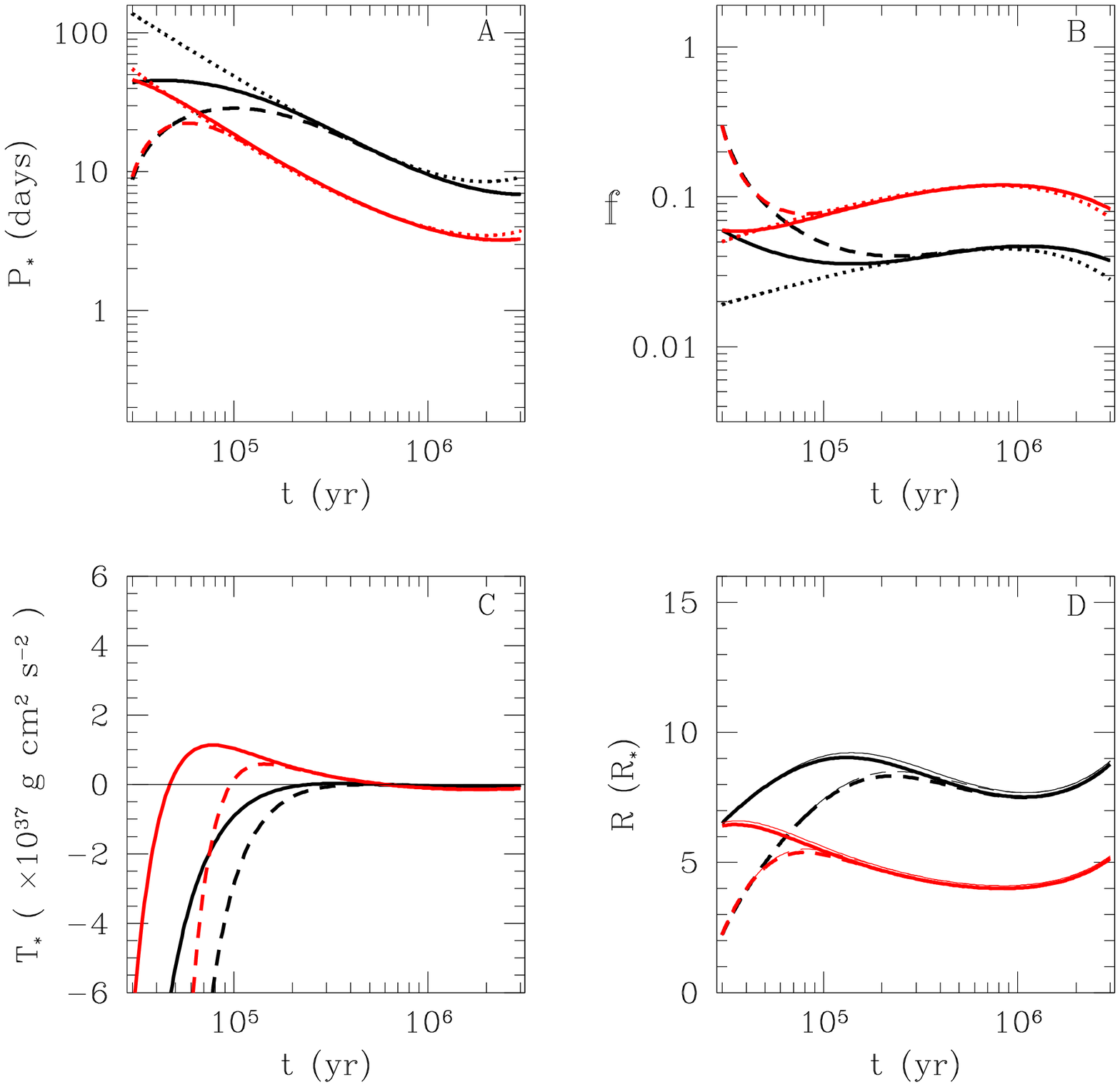}
\caption{Results for $B_*=2000$~G, $\gamma_c=1$, and $\beta=1$, in the
  same format as figure \ref{fig_g1b500bet01}.}
\label{fig_g1b2000bet1}
\end{figure*}

     \subsection{Effect of the Magnetic Star-Disk Interaction}
\label{sec_gam1}

The results in the previous section demonstrate how the spins of
accreting stars are expected to evolve in the absence of magnetic
effects.  In this section, we include the magnetic torques that arise
from the magnetic star-disk interaction.  The torque calculation
follows MP05 (described in \S \ref{sec_sdint}), and we adopt
$\gamma_c=1$ in order to include the effect of the opening of the
field that is expected to arise due to the differential rotation
between the star and disk.

Figure \ref{fig_g1b500bet01} shows the results for cases with a dipole
field strength of $B_*=500$~G at the equator of the star, and
$\beta=0.01$.  This field strength is similar to the strongest yet
measured dipole field on an accreting young star, 600~G for BP Tau
\citep{donatiea08}\footnote{Note that \citet{donatiea08} report the
  polar field strength of 1.2~kG, while $B_*$ represents the
  equatorial value (see \S \ref{sub_magfield}).  For a dipole, $B_*$
  is a factor of two lower than the polar field strength.}.  This
value for $\beta$ was suggested by MP05 to best represent real systems
(see \S \ref{sub_coupling}).  The four models are shown in the same
format as figure \ref{fig_b0} and represent case O1 in table
\ref{tab_parms}.  A comparison of the upper panels of figures
\ref{fig_b0} and \ref{fig_g1b500bet01} reveals that the magnetic
star-disk interaction has very little influence on the spin evolution,
in this case.  In most models, the presence of a 500 G dipole field
results in spin rates that are 10-30\% faster at an age of 3~Myr than
the $B_*=0$ case.  Only the model with low accretion rate and fast
initial spin (black, dashed line) spins more slowly (by $\sim10$\%)
than the non-magnetic case.

Panel (d) of figure \ref{fig_g1b500bet01} indicates that in all
models, the disk is truncated very close to the corotation radius (the
lines for $R_t$ and $R_{\rm co}$ are indistinguishable).  Thus, the
500 G dipole is strong enough to truncate the disk, leading to
magnetospheric accretion, in spite of the fact that the stellar spin
evolution is not largely affected.

The red and black dotted lines in panels (a) and (b) of figure
\ref{fig_g1b500bet01} represent the hypothetical equilibrium (disk
locked) spin rate (see \S \ref{sub_equilib}) for the two different
accretion rates considered.  It is clear that the three models that
spin faster than the non-magnetic case are all spinning more slowly
than their corresponding equilibrium rate.  This means that the net
torque on these stars is positive (acting to spin up the star).
Furthermore, a comparison of panels (c) in figures \ref{fig_b0} and
\ref{fig_g1b500bet01} indicates that these three models have a larger
net torque in the magnetized (O1) case.  This explains why these
models end up spinning slightly faster than the non-magnetized case.
The model with low accretion rate and fast initial spin (black, dashed
line) evolves near the equilibrium rate, resulting in a near zero
torque in panel (c).  However, this is simply due to a coincidence of
the initial condition with the equilibrium rate, since the torque is
not strong enouth to maintain equilibrium within million-year
timescales.

In the disk-locking scenario, the magnetic star-disk interaction
torques are strong enough to keep the spins near their equilibrium
rate.  The case with $B_*=500$~G presented in the appendix (C1, figure
\ref{fig_ginfb500bet1}), where field line opening is ignored, does
approximately follow the expectations of disk locking.  However, this
is clearly not true for the O1 case (figure \ref{fig_g1b500bet01}),
since the spins of all four models at all times depend upon the choice
of initial spin rate.  The main result of case O1 is that, when
considering the opening of the field and strong magnetic coupling, a
dipole magnetic field strength of 500 G is not strong enough to
significantly influence the spin evolution of one solar mass accreting
pre-main-sequence stars.

Figure \ref{fig_g1b2000bet01} shows the results for the O2 case, which
is the same as O1, except that the magnetic field is much stronger
($B_*=2000$~G).  In the O2 case, the magnetic star-disk interaction
has more of an influence on the spin evolution.  By comparing the top
panels of figures \ref{fig_g1b500bet01} and \ref{fig_g1b2000bet01}, it
is clear that all models are spinning more slowly in the O2 case.

The spin rates of the two models with the high accretion rate (red,
solid and red, dashed lines) converge toward the equilibrium value at
an age of $\sim1$~Myr.  Thus, these models approach the disk locked
state, and the initial condition is ``erased'' at that time.  The same
is not true for the models with the low accretion rate (black, solid
and black, dashed lines), since the spin rate after 3~Myr still
depends upon the intial spin rate for those models.

The O2 case is significantly different than the case with $B_*=2000$~G
presented in the appendix (C2, figure \ref{fig_ginfb2000bet1}), which
ignores field line opening.  By comparison, the disk locking in the
O2, high-accretion models is marginal and only happens after
$\sim1$~Myr, which is approaching the lifetime of disks.  Furthermore,
it is important to note that the rotation periods of all models in the
O2 case are less than $\sim3$~days in the age range of 1--3~Myr.  This
is consistent only with the fastest rotators in the observed spin
distributions (see \S \ref{sec_intro}).

Thus far, we have considered $\beta=0.01$, since this was suggested by
MP05.  However, MP05 showed that the spin-down torque felt by the star
is the strongest for a magnetic coupling strength of $\beta=1$.  This
is indeed a highly diffusive situation, as this means that a magnetic
field line that is tilted by 45$^\circ$ will slip through the disk at
a speed equal to the Keplerian speed (see MP05).  Although this is
unexpectedly diffusive for an astrophysical plasma, the correct value
of $\beta$ for these systems is highly uncertain.  Thus, it is
instructive to look at this case, which will have the most extreme
effect on the stellar spins.

Figure \ref{fig_g1b2000bet1} shows the results of this extreme case,
with $\beta=1$ and $B_*=2000$~G (case O3).  It is clear that, in this
case, the star-disk interaction completely dominates the spin
evolution of the stars.  The spin rates converge toward their
respective equilibrium values in 1--3$\times10^5$~yr.  Subsequent
evolution closely follows the equilibria, consistent with the disk
locking scenario.  The spin periods are in the range of 3--10~days for
ages of 1--3~Myr.  It is evident from panel (d) that the disk
truncation radii are close to the corotation radii, but not so close
that the lines overlap.  This difference from the previous 2 cases is
due primarily to the much higher magnetic diffusion rate in the O3
case.

\section{Discussion and Conclusions} \label{sec_conclusion}

The star-disk interaction model of MP05 includes the effects of the
field opening expected to arise from star-disk differential rotation
($\gamma_c=1$), as well as the effects of magnetic coupling,
parameterized by $\beta$.  Based essentially upon the idea that the
disk likely exists in a state of high magnetic Reynolds number, MP05
argued that a value of $\beta=0.01$ is appropriate for real systems.
In the present work, we have extended the MP05 torque model into the
time domain, in order to explore the consequences of magnetic
star-disk coupling and connectedness on the evolution of stellar
spins.

To this end, we developed a stellar spin evolution model (\S
\ref{sec_model}).  The model follows an accreting, one solar mass star
during the Hayashi phase (from $3\times10^4$ yr to $3$ Myr) and
considers the torques expected to arise from the magnetic star-disk
interaction.  The primary goals of this work were to examine the role
of disk locking and to determine if the star-disk interaction torques
alone are sufficient to explain the observed range of spin rates.  In
developing our torque and spin-evolution model, we have adopted many
of the same assumptions as the disk-locking models in the literature,
except that the MP05 torque formulation includes the effect of the
opening of the magnetic field.  Thus, while there are inherant
uncertainties associated with the adoption of many of these
assumptions, our results serve best to highlight the effect of field
opening {\it relative} to previous results in the literature.

Given the expectation that $\gamma_c=1$ and $\beta=0.01$
best-represents the conditions in real systems, and for a range of
accretion rates and field strengths appropriate for T Tauri stars, the
two main conclusions of the present work are as follows:

1. The models presented in figures \ref{fig_g1b500bet01} and
\ref{fig_g1b2000bet01} exhibit spin periods ranging from 3 days to
less than 1/3 day, in the age range of 1--3~Myr (see \S
\ref{sec_gam1}).  These are consistent only with the fastest rotators
in the observed spin period distribution, which is significantly
populated from approximately 1--10~days (see \S \ref{sec_intro}).  It
is apparent that the torque arising from the magnetic connection
between the star and disk is not sufficient to explain the relatively
slowly spinning, young stars.

2. Furthermore, the stars in these models are generally not in spin
equilibrium (with a net zero torque), and the spin rate at all times
depends upon the initial spin rate (see \S \ref{sec_gam1}).  Only the
models with the strongest field strength and highest accretion rate
neared their equilibrium spin rate in $\sim1$~Myr (see figure
\ref{fig_g1b2000bet01}).  However, this equilibrium had a relatively
short spin period of less than a day.  It is apparent that disk
locking does not play a strong role, except possibly for the most
rapid rotators.

Note that the term ``disk locking'' generally refers to the idea that
the angular spin rate of the star is nearly equal to that of the inner
edge of the disk (i.e., $R_t \approx R_{\rm co}$).  In our magnetic
models, this condition was always true.  In fact, the truncation
radius was nearest to the corotation radius in the models that were
furthest from spin equilibrium (e.g., compare panel (d) in figures
\ref{fig_g1b500bet01} and \ref{fig_ginfb500bet1}).  Thus, these stars
do have an angular spin rate that is very close to that of the disk
inner edge, although it is not appropriate to think of these systems
as being ``locked'' to any particular spin rate.  Instead, this is a
consequence of how the disk is truncated.  When the magnetic coupling
in the disk is strong (small $\beta$), the disk will generally be
truncated close to the corotation radius (in State 2; see \S
\ref{sec_rt}).  But, the condition that $R_t \approx R_{\rm co}$ does
not necessarily mean that the star experiences a net zero torque.

In order to explain the existence of slowly rotating young stars, it
is necessary to consider other possibilities.  Since the magnetic
torques depend strongly on the magnetic field strength, it is tempting
to suggest that a stronger magnetic field may solve the problem.
Using our model, we find that (for $\beta=0.01$) a dipole field
strength of $B_*=10^4$~G is required to maintain spin periods
consistent with the slow rotators. However, no T Tauri star has been
found to have surface field strengths greater than 3~kG
\citep{johnskrull07}, and the global (dipole) fields are generally
even weaker \citep[e.g.,][]{safier98, bouvierea07, donatiea07,
  donatiea08}.  Given the relatively small number of stars for which
there are magnetic field measurements, the present situation could be
improved by more and improved observations of the global field
strength and geometry.

Since the models presented in section \ref{sec_gam1} with weak
magnetic coupling (figure \ref{fig_g1b2000bet1}) exhibit slow spins,
it is also appropriate to consider the possibility that real systems
have weak coupling.  As discussed above, the conditions that are
expected to best represent young star-disk systems from ``first
principles'' suggest values of $\beta=0.01$.  However, $\beta$ is a
highly uncertain parameter because the physics of magnetic coupling is
not well understood, nor are the conditions that influence the
coupling (e.g., the ionization fraction of disk gas, turbulence levels
in the disk, or magnetic reconnection rates).  In order for magnetic
coupling to explain the slow rotators, the coupling strength must not
be very different (neither larger nor smaller) than $\beta=1$, since
MP05 showed that the maximum spin-down torque occurs for a $\beta$ of
unity.  The condition that $\beta \approx 1$ seems to require some
fine-tuning, which makes this possiblity even more difficult to
justify at present.

Therefore, it appears necessary to consider effects that are
alternative or additional to the magnetic torques arising from the
star-disk connection.  Real systems seem to be variable on all
timescales \citep[e.g.,][]{hartmann97}, and we have neglected
variability here.  It may be possible, for example, that short
timescale (e.g., $\la 10^4$~yr), large-magnitude variations of the
accretion rate are important for the stellar mass and angular momentum
evolution \citep[e.g.,][]{popham96}.  Finally, it is necessary to
consider the angular momentum loss from pre-main-sequence stellar
winds (see \S \ref{sec_intro}).  The idea that stellar winds may be
important during the accretion phase is well-supported
\citep[e.g.,][]{decampli81, hartmann3ea82, kwantademaru88,
  hartmannea90, fendt3ea95, fendtcamenzind96, safier98,
  bogovalovtsinganos01, sauty3ea02, edwardsea03, edwardsea06,
  dupreeea05, meliani3ea06, kurosawa3ea06, kwan3ea07, fendt09}.
Stellar winds may be powered by the accretion process itself
\citep{toutpringle92, cranmer08} and be the key driver of angular
momentum loss \citep{hartmannmacgregor82, mestel84,
  hartmannstauffer89, paatzcamenzind96, mattpudritz05l,
  mattpudritz08II, mattpudritz08III}.

Some recent magnetohydrodynamic (MHD) simulations of rapidly rotating
stars \citep[e.g.,][]{romanovaea05, romanovaea09} exhibit a type of
``propeller'' regime in which intermittent accretion occurs on typical
timescales of several orbits of the inner disk, while most of the
would-be accreting material is instead launched into a wind.  For the
calculations in the present work, the torque represents that which is
averaged over a calculation timestep.  Our timesteps range from
approximately 400 years (at the earliest times) to $\sim 10^5$ yrs.
While some of the simulations in the work cited above have been run
for thousands of dynamical times, this is still orders of magnitude
shorter than the typical timestep in the present work.  It is not
clear whether the torque formulation we adopted accurately describes
the time-average behavior of the accretion and closed field region in
the propeller regime simulations.  In addition, while these MHD
simulations reliably demonstrate many of the complexities and
sometimes episodic nature of the magnetic star-disk interaction, when
considering the long-term evolution of the star and disk, it is not
yet clear whether the duration of the episodic (e.g., propeller)
regime is significant for the overall spin evolution of young stars.
However, the MHD simulations such as those cited above often exhibit
significant outflows and torques arising from open magnetic field
regions.  In this respect, our result that the torques arising only
from closed field regions are not sufficient to significantly
spin-down the star, agrees with the simulations.

It is clear that the opening of field lines is an important effect,
since including this in the models indicates considerably different
physics at work than the closed field models, when comparing to the
same observational data.  In future work, we will extend the present
analysis to include the angular momentum loss from accretion-powered
stellar winds.

\acknowledgments

The authors are grateful for useful suggestions and comments from
Nairn Baliber and Ralph Pudritz.  SPM was supported by an appointment
to the NASA Postdoctoral Program at Ames Research Center, administered
by Oak Ridge Associated Universities through a contract with NASA.
TPG acknowledges support from NASA's Origins of Solar Systems program
via WBS 811073.02.07.01.89.

\appendix  \label{sec_gaminf}

\section{Cases With Closed Magnetic Field (\mbox{$\gamma_{\lowercase{c}} = \infty$})}

\begin{figure*}
\epsscale{\psiz}
\plotone{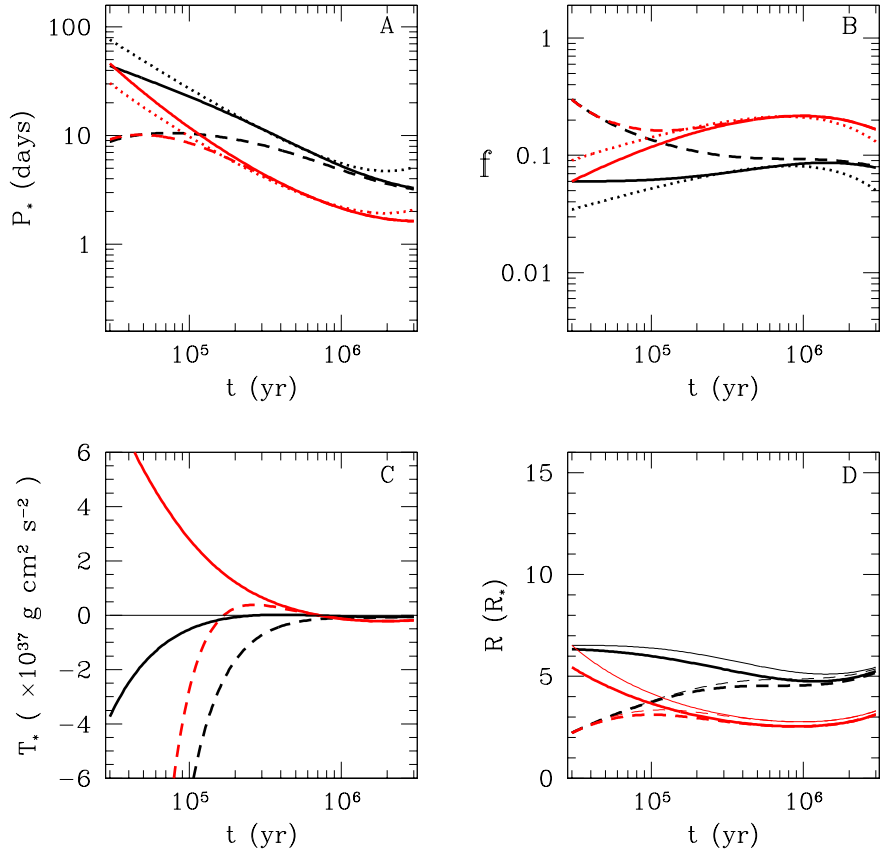}
\caption{Results for $B_*=500$~G, $\gamma_c=\infty$, and $\beta=1$, in
  the same format as figure \ref{fig_g1b500bet01}.}
\label{fig_ginfb500bet1}
\end{figure*}

\begin{figure*}
\epsscale{\psiz}
\plotone{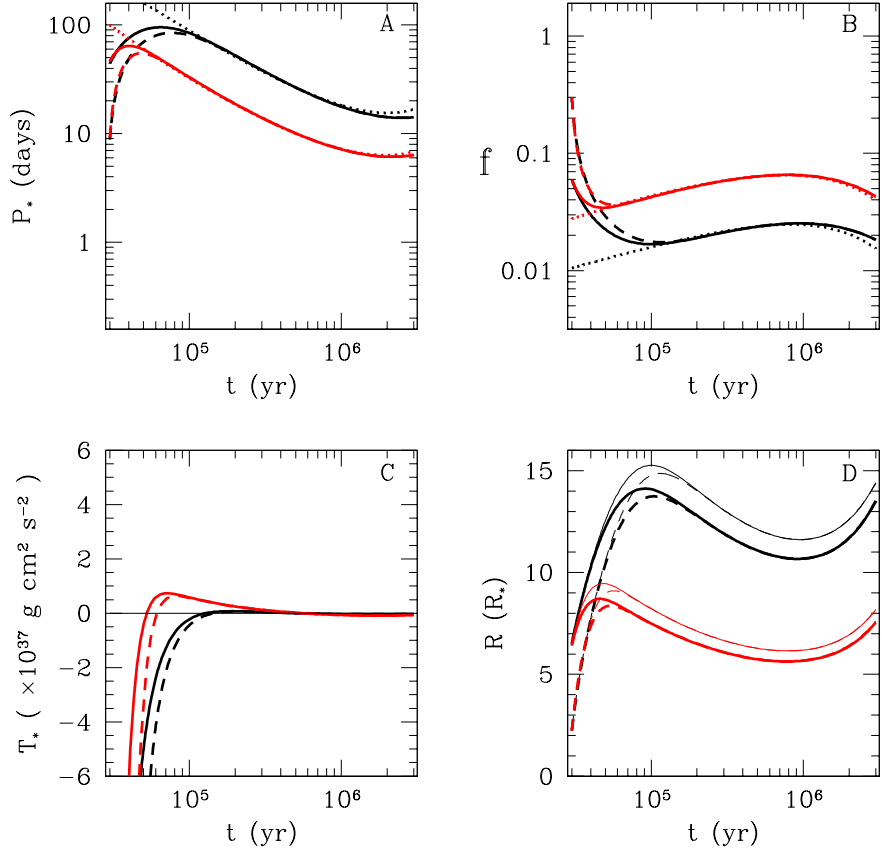}
\caption{Results for $B_*=2000$~G, $\gamma_c=\infty$, and $\beta=1$,
  in the same format as figure \ref{fig_g1b500bet01}.}
\label{fig_ginfb2000bet1}
\end{figure*}

In order to demonstrate that our model and code reproduces the results
of the classical models, and for comparison to the cases presented in
section \ref{sec_results}, here we present cases with
$\gamma_c=\infty$.  With the MP05 torque formulation, this value of
$\gamma_c$ keeps the magnetic field connected everywhere to the disk,
regardless of how twisted the field becomes.  This well-represents the
classical \citet{ghoshlamb78} model, and results in stronger spin-down
torques on the star.  Most star-disk interaction and disk-locking
models follow the Ghosh \& Lamb model and assume a magnetic coupling
strength that is equivalent (within $\sim$20\%) to a value of
$\beta=1$.  Thus, we adopt this value of $\beta$ here.

In table \ref{tab_parms}, the two cases presented here are labeled C1
and C2, which have $B_*=500$~G and $B_*=2000$~G, respectively.
Figures \ref{fig_ginfb500bet1} and \ref{fig_ginfb2000bet1} show
results from these two cases, where the line styles and colors have
the same meaning as in figure \ref{fig_g1b500bet01}.  The spin
evolution of all models in both figures is dominated by the magnetic
interaction between the star and disk.  For the $B_*=500$~G case
(figure \ref{fig_ginfb500bet1}), the spin rates converge toward their
equilibrium values in $2\times10^5$ and $10^6$~yr for the cases with
high and low accretion rates, respectively.  For the $B_*=2000$~G case
(figure \ref{fig_ginfb2000bet1}), convergence occurs for all models
within $\sim10^5$~yr.  Thus, when the magnetic field is not allowed to
open as expected, the evolution follows closely with the expectations
of the disk locking scenario.

As in all other magnetic models presented in this paper, the disk
truncation radii remain near, although slightly inside, the corotation
radii.  However, in these cases with $\gamma_c=\infty$, $R_t$ is
generally further away from $R_{\rm co}$ than in the cases presented
in section \ref{sec_results}.  This is a consequence of the stronger
spin-down torques and higher value of $\beta$ for the models of this
section.

Examination of figures \ref{fig_ginfb500bet1} and
\ref{fig_ginfb2000bet1} reveals that the spin rates are slower when
the magnetic field strength is higher or the accretion rate is lower.
Under the conditions of disk locking, the angular spin rate of the
star is proportional to $\dot M_a^{3/7}B_*^{-6/7}$, which is exhibited
in the models presented here.  Furthermore, the spin periods of all
the models are in the range of 1.5--20~days during the ages of
1--3~Myr.  These values are consistent with the range of observed spin
periods (see \S \ref{sec_intro}), and this has been the primary
physical justification for the disk-locking picture being applied to
young stars.

The models presented in this section produce very similar quantitative
results to the spin evolution models of \citet{yi94, yi95} and AC96.
The main differences between all of these works is the treatment of
the magnetic field and accretion rates.  In particular, other works
considered a magnetic field strength that evolved with stellar spin
rate and stellar radius.  Here, we have assumed a constant surface
magnetic field strength.  Our assumption results in different behavior
from those models, particularly at times earlier than $\sim1$~Myr.  At
these younger ages, the stars have relatively large radii, and the
assumption of a constant surface field strength means that the total
magnetic energy in the field (proportional to $B_*^2R_*^3$) is larger.
This is the main reason that the equilibrium spin rates are slowest at
earlier times, in our models.  The models of \citet{yi94, yi95} and
AC96 exhibit spin periods that are more nearly constant in time, due
the different prescription of the field.  Still, when our results and
those of \citet{yi94, yi95} and AC96 are scaled according to the
torque theory to similar values of accretion rates and field
strengths, the results are quantitatively consistent.

The spin evolution model of CC93 is somewhat different from other
works.  The assumptions in that work lead to a magnetic coupling that
is significantly stronger than in \citet{yi94, yi95}, AC96, and most
other models.  We have determined that we get similar results to CC93
for models with $\gamma_c=\infty$ and a coupling strength of
$\beta=0.05$.  This relatively strong coupling leads to a large
azimuthal twisting of the magnetic field.  This, together with the
requirement that the magnetic field remain closed, leads to very
strong spin-down torques on the star.  For this reason, the CC93 model
predicts slower equilibrium spins than other models for a given field
strength \citep[e.g., see][]{johnskrull3ea99, johnskrull07}.  However,
as shown in section \ref{sec_results}, when the opening of the field
is properly taken into account, the spin-down torques are weaker, and
the magnitude of this effect is larger for cases with strong magnetic
coupling (as in the CC93 model).



\end{document}